\documentstyle [12pt]{article}

\catcode`\@=11
\def\makepreprititle{\par
  \begingroup
  \def\thefootnote{\fnsymbol{footnote}}
  \def\
@makefnmark{\hbox
  to 0pt{$^{\@thefnmark}$\hss}}
  \if@twocolumn
  \twocolumn[\@makepreprititle]
  \else \newpage
  \global\@topnum\z@
  \@makepreprititle \fi\thispagestyle{empty}\@thanks
  \endgroup
  \setcounter{footnote}{0}
  \let\makepreprititle\relax
  \let\@makepreprititle\relax
  \gdef\@thanks{}\gdef\@author{}\gdef\@title{}
  \gdef\@preprintnumber{}\gdef\@preprintdate{}\gdef\subtitle{}
  \let\thanks\relax}
\def\preprintnumber#1{\gdef\@preprintnumber{#1}}
\def\preprintdate#1{\gdef\@preprintdate{#1}}
\def\subtitle#1{\gdef\@subtitle{#1}}
\def\@makepreprititle{\newpage
{\def\baselinestretch{1}
  \begin{flushright} \@preprintnumber \par
  \@preprintdate \end{flushright} } \par
  \begin{center}
\vskip 1.5em
  {\LARGE \@title \par} \vskip 2.5em
  {\Large \lineskip .5em
  \begin{tabular}[t]{c}\@author
  \end{tabular}\par}
  \vskip 1em {\large \@date} \end{center}
  \par
  \vfil}
\date{\sl Department of Physics, Tohoku University\\Sendai, 980 Japan}
\preprintdate{}
\preprintnumber{}
\subtitle{}
\def\abstract{\if@twocolumn
\section*{Abstract}
\else \normalsize
\begin{center}
{\bf Abstract\vspace{-.5em}\vspace{0pt}}
\end{center}
\quotation
\addtocounter{page}{-1}
\fi}
\def\endabstract{\if@twocolumn\else\endquotation\fi}
\catcode`\@=12
\def\thebibliography#1{\section*
 {References					
 \markboth{REFERENCES}{REFERENCES}}\list
 {[\arabic{enumi}]}				
 {\settowidth\labelwidth{[#1]}\leftmargin\labelwidth
 \advance\leftmargin\labelsep
 \usecounter{enumi}}
 \def\newblock{\hskip .11em plus .33em minus -.07em}
 \sloppy \sfcode`\.=1000\relax}

\renewcommand{\thefootnote}{\fnsymbol{footnote}}
\newcommand{\bea}{\begin{eqnarray}}
\newcommand{\eea}{\end{eqnarray}}
\newcommand{\simgt}{\hbox{ \raise3pt\hbox to 0pt{$>$}\raise-3pt\hbox{$\sim$} }}
\newcommand{\simlt}{\hbox{ \raise3pt\hbox to 0pt{$<$}\raise-3pt\hbox{$\sim$} }}
\newcommand \vc[1]{{\bf {#1}}}

\def\to{\rightarrow}
\def\epem{\ifmmode{ e^{+}e^-} \else{$ e^{+}e^- $ } \fi}
\def\bw{\ifmmode{ bW^+ } \else{$ bW^+ $ } \fi}
\def\bwb{\ifmmode{ \bar{b}W^- } \else{$ \bar{b}W^- $ } \fi}
\def\ttbar{\ifmmode{t\bar{t}} \else{$t\bar{t}$} \fi}
\def\nrg{\ifmmode{\tilde{G}(\vc{p};E)} \else{$\tilde{G}(\vc{p};E)$} \fi}

\def\alpsmz{\alpha_{s}(m_Z)_{\overline{\bf MS}}}

\def \lsa {\rlap {\lower 3.5 pt \hbox {$\mathchar \sim$}} \raise 1
pt \hbox {$<$}}
\def \rsa {\rlap {\lower 3.5 pt \hbox {$\mathchar \sim$}} \raise 1
pt \hbox {$>$}}
\def\msbar {\ifmmode{\overline{\rm MS}}    \else{$\overline{\rm MS}$ }    \fi}
\def\oalfs {$O(\alpha_s)$}

\def\alpsmz{\ifmmode{\alpha_s(m_Z)_{\overline{\rm MS}}}
              \else{$\alpha_s(m_Z)_{\overline{\rm MS}}$} \fi}
\def\lamfive{\ifmmode{\Lambda^{(5)}_{\overline{\rm MS}}}
               \else{$\Lambda^{(5)}_{\overline{\rm MS}}$} \fi}
\def\lamfour{\ifmmode{\Lambda^{(4)}_{\overline{\rm MS}}}
               \else{$\Lambda^{(4)}_{\overline{\rm MS}}$} \fi}
\def\lsa{\rlap{\lower 3.5 pt \hbox {$\mathchar \sim$}} \raise 1pt \hbox {$<$}}
\def\rsa{\rlap{\lower 3.5 pt \hbox {$\mathchar \sim$}} \raise 1pt \hbox {$>$}}

\begin{document}


\preprintnumber{TU--469}
\preprintdate{October, 1994}

\title{
Review on Physics of $e^+e^- \to t\bar{t}$ Near Threshold$^*$
}

\author{Y.~Sumino}
\date{\sl  Department of Physics, Tohoku University, Sendai, 980 Japan}
\makepreprititle

\begin{abstract}
\normalsize

We review the recent developments of studies on
$e^+ e^- \to t\bar{t}$ in the threshold region.
It has been shown that the top threshold cross sections will be able to
provide many important and interesting physics.
Due to the large mass and the large decay width of top quark,
there will appear a number of unique features at top threshold.
First, theoretical background is summarized.
Then we present observables
that are typical to the top threshold region, and
discuss the physics that can be extracted.
Also, we list the results of the recent parameter determination study
taking a sample case of $m_t = 170$~GeV.

\end{abstract}
\vspace{1cm}

\vfil

\footnoterule
\footnotesize
\vspace{3mm}
\hspace{-6mm}
$^*$ Talk given at XXXIV Cracow School of Theoretical Physics, Zakopane,
Poland, June, 1994.

\newpage

\normalsize
\section{Introduction}

We review the physics of
$\epem \to \ttbar$ near threshold ($E_{c.m.} \simeq 2m_t$)
in this paper.
In the beginning
let us briefly comment on the role of
future \epem colliders in order to
give some idea to the readers on the status of
future top quark
physics from a more general viewpoint.

\subsection{About NLC}

It has recently been recognized that a next generation linear \epem
collider (NLC), with energy accessible up to $\sim 500$~GeV in its
first stage operation, will be able to provide important contributions
to future elementary particle physics field.\cite{nlc}
Possibility to build such a collider is now seriously considered at
KEK, DESY, and SLAC.
These machines aim at the luminosity of $10^{33}$-$10^{34} ~
\mbox{cm}^{-2}
\mbox{sec}^{-1}$,
which corresponds to $10^5$-$10^6$ $W$'s and $10^3$-$10^4$
top quarks produced per year assuming the standard model
cross sections.
Also, due to the characteristics of linear colliders, the beam energy
will be adjustable downward within a range of
a few hundred GeV, a highly
longitudinally polarized electron beam can be used, etc.

NLC will be designed so that it can test the standard model predictions
with good accuracy at the energy approximating the
electroweak symmetry breaking scale.
Therefore,
one may address the aim of NLC as the investigation of electroweak
symmetry breaking physics and the search for any new physics to appear.
Target particles necessarily are
$W$ and $Z$, Higgs particle and top quark, and not to mention
non-standard-model
particles if any.
As the $W$ and $Z$ are massive, we know their longitudinal components come
from the electroweak
symmetry breaking sector, and we may extract information
on it from detailed studies of the properties of the gauge bosons.
The Higgs is the very particle that is responsible for the symmetry
breaking mechanism within the standard model (and also in many of its
extensions) so that its discovery and/or the
detailed study of its properties will be crucial in understanding the
fundamental physics.
The top quark is the heaviest fermion in the standard model with the
mass of the order of the electroweak symmetry breaking scale, and one
may again expect to probe the symmetry breaking physics using top quark.

In the rest of this
paper we focus on the physics that can be extracted
through the study of top quark pair production process near
threshold at NLC.
It turns out that the study of top quark threshold is quite promising and
also very interesting among the various subjects of NLC.
The threshold cross section depends on physical parameters such
as top mass $m_t$,
strong coupling constant $\alpha_s$, top width $\Gamma_t$, Higgs mass
$M_H$, top-Higgs Yukawa coupling $g_{tH}$, etc., as we will see in the
following.

\subsection{What Do We Expect at Top Threshold?}

As compared to all other fermions,
two of the unique properties of top quark are
that it is extremely heavy and that it has very short lifetime.
CDF group has recently announced the evidence for \ttbar
production\cite{cdf},
and the reported
top quark mass
\bea
m_t = 174 \pm 10
\begin{array}{l} +13 \\ -12 \end{array}
{}~\mbox{GeV}
\eea
agrees very well with the indirect determination from the LEP and
SLAC data.\cite{hmhk}
The top quark decays almost 100\% to
$b$ quark and $W$
within the standard model.
The decay width of top quark $\Gamma_t$ is predictable
as a function of $m_t$, and
already fairly precise theoretical prediction
at the level of a few percent accuracy is available.\cite{jk1}
Here, we note that $\Gamma_t \sim 1$~GeV for the
relevant top quark mass.

Near the threshold of top quark pair production one might expect the
formation of \ttbar resonances.
Along with the resonance formation,
cross section gets enhanced by QCD interaction.
Produced $t$ and $\bar{t}$, however, decay quickly via electroweak
interaction and the enhancement is reduced accordingly.
In fact there will appear subtle interplays between QCD
and electroweak interactions.

Consider the time evolution of
the \ttbar pair produced in $\epem$
annihilation as they spread
apart from each other.\ (Fig.~1) \,
Since they are slow near the threshold, they cannot
escape
even relatively weak attractive force mediated by the exchange of
Coulomb gluons;
$t$ and $\bar{t}$ are bound to form the Coulombic resonances
when they reach the distance of Bohr radius
$(\alpha_s m_t)^{-1} \sim 0.1$~GeV$^{-1}$.
At this stage, the coupling of top quark to gluon
is of the order of $\alpha_s(\mu \! = \! \alpha_sm_t) \sim 0.15$.
If they could continue to
spread apart even further to the distance
$\Lambda_{QCD}^{-1} \sim$ a few GeV$^{-1}$,
there would occur the hadronization
effect as the coupling becomes really strong,
since gluons with wave-length $\sim \Lambda_{QCD}^{-1}$ would be able to
resolve
the color charge of each constituent.
For a realistic top quark, however, the \ttbar pair will
decay at the distance $(m_t\Gamma_t)^{-1/2} \sim 0.1$~GeV$^{-1}$
into energetic $b$ and $\bar{b}$ jets and $W$'s
before the hadronization
effect becomes important.
Here, the toponium can be regarded as the Coulombic resonance
state (with reasonably weak coupling)
due to the large mass and width of top quark.
In this respect, the toponium resonances differ distinctly
from the charmonium and bottomonium
resonances
which have smaller masses as well as narrow widths.
Besides, since the toponium resonances decay
dominantly via electroweak
interaction\cite{hkmn,kz},
their decay process can also be calculated reliably.

The distinct
resonance shape of the total cross section that would appear for
narrow resonances gets smeared due to large top quark width,
merging into a broad
enhancement of the cross section over the threshold region.
(See Fig.~6.a.)
Generally, the QCD enhancement is still large enough to allow precise
test of QCD in the \ttbar threshold region,
but one is obliged to study the overall shape of the
total cross section
instead of applying the spectroscopic method developed at
charmonium and bottomonium resonances.\cite{fk,sp}
Besides total cross section there are two other observables
that are unique to the \ttbar threshold region.
It was pointed out that through
the measurement of top quark momentum distribution in the
threshold region
one can extract information on the QCD binding effect independent of that
from the total cross section.\cite{jlc,jkt}
Also there will appear observable forward-backward (FB)
asymmetry of top quark even below threshold,
which provides
yet another independent information of the toponium resonances.\cite{fb}

In section 2, we review the basic theoretical aspects of \ttbar
threshold which allow for reliable estimates of cross sections.
In section 3, we present various observables at \ttbar threshold, and
discuss the physics that can be extracted.
Section 4 lists the results of the recent parameter determination study
from the \ttbar threshold cross sections.
Concluding remarks are given in section 5.

\section{Theoretical Background}

Fadin and Khoze first pointed out that the top quark threshold
cross section will provide
a very clean test of QCD since fairly stable theoretical prediction
is available.\cite{fk}
First, the large top quark mass enables the probe of
asymptotic region ($\mu \simgt 10$~GeV)
of QCD where the theoretical control is more feasible.
Secondly, the large width of top quark will act as the infra-red cutoff
which prevents the hadronization effects to affect the cross section.
Thirdly, spacelike region of gluon momentum
plays the prime role in the
QCD enhancement.

In this section we will review the basic
theoretical concept necessary for
dealing with the threshold bound states.
We see that non-perturbative yet systematic discussion is possible.
The upshot is that the \ttbar production vertex can be calculated
in terms of the Green's
function of the non-relativistic Schr\"{o}dinger equation.\cite{fk}

\subsection{Leading Threshold Singularities}

It is well-known that
near the threshold of quark-antiquark ($q\bar{q}$) pair
production, naive perturbation theory breaks
down due to the formation of bound states \cite{braun,ap}.
Intuitively, this is because the produced $q$ and $\bar{q}$
have small velocities so that they are trapped by the
attractive force mediated by the exchange of
gluons.
Thus, multiple exchange of gluons becomes more significant
and the strong interaction is enhanced accordingly.
We will briefly review this property
near \ttbar threshold.
We demonstrate explicitly that the ladder diagrams exhibit the
gauge-invariant leading singularities.
In this subsection, we neglect the decay of $t$ and $\bar{t}$,
and treat them as stable particles.
\medbreak

Let us consider the amplitude where a virtual photon
decays into $t$ and $\bar{t}$,
$\gamma^* \to t\bar{t}$, just above the threshold of
$t\bar{t}$ pair production.
As we will see below, the ladder diagram for this
process where uncrossed
gluons are exchanged $n$ times between $t$ and $\bar{t}$
has the behavior $\sim (\alpha_s/\beta)^n$, see Fig.~2.
Here, $\beta$ is
the velocity of $t$ or $\bar{t}$ in the c.m.\ frame,
\bea
\beta = \sqrt{1-\frac{4m_t^2}{s}} .
\label{beta}
\eea
Hence, the contribution of the $n$-th ladder diagram
will not be small even for large $n$ if $\beta \simlt \alpha_s$.
That is,
the higher order terms in $\alpha_s$ remain unsuppressed
in the threshold region.
The singularities which appear at this specific kinematical
configuration is known as ``threshold
singularities''.\footnote{
This singularity stems from the fact that, for a particular
assignment of the loop momenta, all the internal particles
can become very nearly on-shell simultaneously as $\beta \to 0$.
}

We may observe the appearance of $\sim (\alpha_s/\beta )^n$
in the $n$-th ladder diagram as follows.

First, consider the one-loop diagram.
Its {\it imaginary} part can be estimated using the Cutkosky rule
(cut-diagram method), see Fig.~3.
Namely, the imaginary part is
given by the phase space integration of the product of the
tree diagrams.
The intermediate $t\bar{t}$ phase space is proportional to $\beta$ as
\bea
d\Phi_2(t\bar{t}) = \frac{\beta}{16\pi} \, d \cos \theta ,
\eea
where $\theta$ is the angle between the momenta of
the intermediate top quark and the final top
quark in the c.m.\ frame.
Meanwhile the \ttbar scattering diagram with
$t$-channel gluon exchange
contributes the factor $\sim \alpha_s/\beta^2$ since the
gluon propagator is proportional to $1/\beta^2$.
In fact the propagator denominator is given by
\bea
k^2 =
-{\bf k}^2 = - \, \frac{s \beta^2}{2} (1-\cos \theta),
\label{glmomsq}
\eea
where $k \propto \beta$ denotes the gluon momentum.

Thus, we see that the imaginary part of the one-loop diagram
has the behavior $\sim \beta \times \alpha_s /\beta^2 = \alpha_s/\beta$.
By repeatedly using the cut-diagram method, one can induce
that the imaginary part of the
ladder diagram with $n$ uncrossed gluons has the
behavior $\sim (\alpha_s/\beta)^n$, see Fig.~4.

Note that the leading part
of the gluon propagator in powers of $\beta$
(in $R_\xi$-gauge)
is the instantaneous (Coulomb) propagator as
\bea
\bar{u}_f \gamma^\mu u_I \, \,
\frac{-i}{k^2+i\epsilon}
\left[
g_{\mu \nu} - ( 1 \! - \! \xi ) \frac{k_\mu k_\nu}{k^2}
\right]
\bar{v}_I \gamma^\nu v_f
\nonumber \\
{}~~ \to ~~
\bar{u}_f \gamma^0 u_I \, \,
\frac{ -i }{-{\bf k}^2 + i \epsilon}
\, \,
\bar{v}_I \gamma^0 v_f .
\label{glp}
\eea
Here, the subscripts $f$ and $I$ stand for the final state and
the intermediate state, respectively.
We used the fact that the space components of the currents,
$\bar{u}_f \gamma^\mu u_I$ and $\bar{v}_I \gamma^\nu v_f$,
are order $\beta$ in the c.m.\ frame.\footnote{
Dirac representation of the $\gamma$-matrices is most useful
in understanding the power count, where $\gamma^0$ is diagonal and
$\gamma^i$'s are off-diagonal.
The $t$-quark
spinor wave function has the upper two components with order
of unity and the lower two components suppressed by $\beta$,
and vice versa for the $\bar{t}$-quark.
}

It can be checked by power counting method\cite{by}
that the {\it real} part of the $n$-th ladder diagram
exhibits the same type of singularity,
$\sim (\alpha_s/\beta)^n$.
The relevant loop momenta in the loop integrals are also
in the non-relativistic regime:
\bea
p_t^0 - m_t, \, \bar{p}_t^0-m_t \sim O(\beta^2),
&&~~
{\bf p}_t = -\bar{{\bf p}}_t \sim O(\beta),
\label{nrmom1}
\\
k^0 \sim O(\beta^2), ~~
&&
{\bf k} \sim O(\beta) .
\label{nrmom2}
\eea
Here, $p_t$, $\bar{p}_t$ and $k$ represent the internal momenta of
$t$, $\bar{t}$ and the gluon, respectively, in the c.m.\ frame.
It is easy to see that,
for such configurations, $t(\bar{t})$ and gluon
propagators are counted as $\sim 1/\beta^2$, and the
measure for the each loop integration $d^4k/(2\pi)^4$ as
$\sim \beta^5$.

Thus, the ladder diagrams exhibit the leading singularities
$\sim (\alpha_s/\beta)^n$.
Other diagrams, including crossed gluon diagrams, do not exhibit
the leading singularities, but contribute to the non-leading
singularities $\sim \alpha_s^{n+l}/\beta^n~$ ($l \geq 1$).

One may worry about the gauge invariance of the amplitude if we take only the
ladder diagrams.
Let us write the amplitude for the process
$e^+e^- \to t\bar{t}$
with full QCD corrections near threshold as
\bea
M^{(full)} (\alpha_s,\beta)
= \sum_n \, c_n (\alpha_s/\beta)^n
\, + \, (\mbox{non-leading terms}).
\eea
Then the coefficients $c_n$'s should be gauge-independent since
the full QCD amplitude $M^{(full)}$
is gauge invariant.
(We suppressed the variables other than $\alpha_s$ and
$\beta$.)
Explicitly, the gauge-independence of $c_n$'s is ensured
by the gauge-independence of the leading part $\sim 1/\beta^2$
of the gluon propagator in eq.~(\ref{glp}).

As the higher order terms in $\alpha_s$ can no longer be neglected
near threshold, we are led to sum over the leading threshold
singularities.
Let us denote by $\Gamma^\mu_0$ the leading singularities of the
vertex $\gamma^* \to t\bar{t}$, which satisfies the self-consistent
equation as depicted in Fig.~5.
By taking only the leading part $\sim (\alpha_s/\beta)^n$ on
both sides of the equation, one obtains\footnote{
See, for example, Ref.\cite{sp} for a fairly systematic derivation of
the vertex $\Gamma^\mu_0$.
}
the
vertex $\Gamma^\mu_0$ as
\bea
\Gamma^\mu_0 = - \,
\left( \frac{1+\gamma^0}{2} \gamma^\mu \frac{1-\gamma^0}{2} \right)
\, ( E - {\bf p}_t^2/m_t  ) \,
\tilde{G_0}( {\bf p}_t;E ),
\label{ttvtx0}
\eea
where $E = \sqrt{s}-2m_t$ is the energy measured from the threshold.
$\tilde{G_0}( {\bf p};E )$ is the S-wave Green's function
of the non-relativistic Schr\"{o}dinger equation with
Coulomb potential:
\bea
&&
\left[
\left( - \, \frac{\nabla^2}{m_t} + V(r) \right)
- ( E + i\epsilon )
\right]
G_0({\bf x};E) = \delta^3({\bf x}),
\label{scheq0}
\\ && ~~~
\tilde{G_0}( {\bf p};E )
= \int d^3{\bf x} \, e^{-i{\bf p} \cdot {\bf x} }
\, G_0({\bf x};E) ,
\\ && ~~~~~~
V(r) = - \, C_F \, \frac{\alpha_s}{r} ,
\eea
where $C_F = 4/3$ is the color factor.
Explicitly, we may write
\bea
\tilde{G}_0(\vc{p};E) = - \sum_n
\frac{\phi_n (\vc{p})\psi_n^* (\vc{0})}{E-E_n+i\epsilon},
\label{swgf0}
\eea
where $\phi_n (\vc{p})$ and $\psi_n (\vc{x})$ are the
Coulomb wave functions in momentum space and coordinate space,
respectively.
Here, $n$ includes the bound states, $E_n=-(C_F\alpha_s)^2 m_t/4n^2$,
and continuum states for $E_n > 0$.
Only the S-wave states contribute to the leading vertex
as seen from the appearance
of $\psi_n(\vc{0})$.\footnote{
To see that
$(E-\vc{p}^2/m_t)\tilde{G_0}({\bf p};E)$ is a function of $\alpha_s/\beta $,
one should identify $E \to m_t \beta^2$ and $|{\bf p}| \to m_t \beta$
in the leading order.
}

As for the amplitude for $e^+e^- \to t\bar{t}$ near threshold,
one may take the leading singularities
$\Gamma^\mu_0 = \sum c^{(0)}_n (\alpha_s^n /\beta^n)$ as the
zeroth order of the new perturbative expansion,
and consider
$\Gamma^\mu_1 = \sum c^{(1)}_n (\alpha_s^{n+1}/\beta^n)$,
$\Gamma^\mu_2 = \sum c^{(2)}_n (\alpha_s^{n+2}/\beta^n)$,
$...$, as the higher order corrections.
They are gauge invariant at each order.
One may regard $\Gamma^\mu_1$ as $O(\alpha_s)$ or $O(\beta)$
correction to the zeroth order $\Gamma^\mu_0$, since
$\alpha_s^{n+1}/\beta^n = \alpha_s (\alpha_s/\beta)^n
= \beta (\alpha_s/\beta)^{n+1}$.\footnote{
More rigorously, one should reorganize the perturbative expansion so as
to include higher order corrections to the resonance mass, the residue
of the resonance pole, etc.
See ref.\cite{nlo} for the detail.
}
Note that the expansion parameter $\beta$ is guaranteed to be
small if $\alpha_s$ is small, since we are interested in the
summation of the leading singularities only in the kinematical
region where the naive perturbation theory breaks down
$(\beta \simlt \alpha_s)$.

Near the \ttbar threshold, $\alpha_s$ will be order 10\% ,
and the above new expansion will be justifiable.
It is important to include the
$O(\alpha_s) = O(\beta )$ corrections to the cross sections
for practical purposes.
(See subsection 3.e.)

\subsection{Top Width As IR Cutoff}

The discussion of the leading threshold singularities in the
previous subsection is identical to that of QED bound states
such as positronium.
In QCD, however, one would expect additional non-perturbative effects
arising from the infra-red region of gluon momentum.
As seen in eq.~(\ref{glmomsq}), the typical gluon momentum scale involved
in the formation of bound states is $k^2 \sim -2m_t^2\beta^2$.
It may even become zero at the threshold, $s=4m_t^2$.

One can show
using the spectral representation of top propagator
that, if we include the large decay width of top quark,
the gluon momentum scale will be effectively cut off as
\bea
| k^2 | \simgt 2 m_t\Gamma_t ,
\eea
and the singular behavior in the amplitude is avoided.\cite{nlo}
At the leading order approximation, the effect of top quark width is
accounted for by replacing $E \to E + i \Gamma_t$ in
eqs.~(\ref{ttvtx0}) and (\ref{scheq0}).\cite{fk}
For instance,
$\sqrt{2m_t\Gamma_t} \simeq 24$ GeV for $m_t=180$ GeV.
(See subsection 1.b.)

\section{Cross Sections Near \ttbar Threshold}

In this section analyses of the cross sections in
the \ttbar threshold region are presented.
There are three independent observables typical to the top quark
threshold, namely, total cross section, top momentum distribution and
FB-asymmetry of top quark.
We discuss the physics that can be extracted from these observables.
Also, the effect of higher-order corrections typical to top threshold is
discussed.

\subsection{Total Cross Section}

The total cross section in the \ttbar threshold region can be obtained
from the leading
\ttbar production vertex (\ref{ttvtx0}) and its higher order
correction.
Using the optical theorem, one finds
\bea
\sigma_{tot}
&=& \sigma_{t\bar{t}} ( \sqrt{s}, m_t, \alpha_s, \Gamma_t, M_H, g_{tH})
\nonumber
\\
&\propto& Im \, G(\vc{x}=0;E)
\simeq - Im \, \sum_n \frac{|\psi_n(\vc{0})|^2}{E-E_n+i\Gamma_t} .
\label{ot}
\eea
It is dependent on various physical
parameters.\cite{fk,sp}\footnote{
Although $\Gamma_t$ can be calculated from $m_t$ in the standard model,
this relation deviates with simple extensions of the model.
e.g.\ $\Gamma_t \propto |V_{tb}|^2$
can be smaller if there is a fourth generation, whereas it
will be greater if there are additional decay modes such as $t \to b H^+$
or $t \to \tilde{t} \tilde{\chi}$.
Thus, it is important to measure $\Gamma_t$.
}
The denominator shows the resonance structure, but due to the large
width $\Gamma_t$, the detail of the structure is smeared out.
The normalization of the cross section is determined by the wave
function at the origin, which is affected by the QCD binding effect.

A sample analysis is given in Fig.~6.
Three lines in Fig~6.a correspond to different values of
$\alpha_s(m_Z)$.
As $\alpha_s$ is increased the shoulder of the cross section moves to
the left due to the increase of the binding energy.
The normalization of the cross section is enhanced at the same time.
{}From the fit of the theoretical prediction to the Monte Carlo
events, $\alpha_s(m_Z)$ and $m_t$
are determined in Fig~6.b.

\subsection{Top Momentum Distribution}

Besides total cross section, one can measure the momentum distribution
of top quark in the threshold region by reconstructing the top quark
momentum $\vc{p}$ from the final $bW$ jet momenta\cite{fms}.
Both 6-jet mode and lepton-plus-4-jet mode can be used to reconstruct
the top momentum.
The top quark momentum distribution in the leading order
is proportional to the square of
the Green's function\cite{jlc,jkt},
\bea
\frac{d\sigma}{d|\vc{p}|} &\propto&
| \tilde{G} ( \vc{p}; E ) |^2
\nonumber
\\
&\simeq &
\biggl| \sum_n
\frac{\phi_n (\vc{p})\psi_n^* (\vc{0})}{E-E_n+i\Gamma_t}
\biggl|^2 .
\label{msgf}
\eea
Therefore, we may measure the momentum space wave functions
of toponium resonances using top quark momentum distribution.
Thus,
the momentum distribution provides information independent of that
from the total cross section, so that it can be used to measure physical
parameters ($\alpha_s, m_t, |V_{tb}|^2$, etc.)\ efficiently
near threshold.
Fig.~7 shows the top momentum distribution for three values of
$\alpha_s(m_Z)$.
The peak position of the distribution turns out to be a most suitable
observable.\cite{fms}

For instance,
the errors on $\alpha_s(m_Z)$ and $m_t$ (statistical only)
are estimated to
be reduced to about 60\% through the simultaneous measurements of the
total cross section and momentum distribution as compared to those
determined from the total cross section alone.\cite{immo}

\subsection{Higgs Effects}

Perhaps Higgs effects are especially interesting.

If the Higgs particle turns out to be light ($M_H \simlt 100$~GeV),
there is a chance to extract the Yukawa coupling of top quark.
Formalism for including Higgs effect has been developed in Ref.\cite{jk2}.
Qualitatively, the effect of Higgs exchange between $t$ and $\bar{t}$
can be understood as including the Yukawa potential
\bea
V_{\mbox{\scriptsize Yuk}}(r)=- \, \frac{g_{tH}^2}{4\pi} \,
\frac{e^{-M_H r}}{r}
\eea
into the
Schr\"{o}dinger equation (\ref{scheq0}).
The effect is larger for smaller $M_H$ and larger $g_{tH}$.
Note that the Higgs mass should not be taken as a parameter, since if the
Higgs effect is of observable size, it will already have been found at
NLC and $M_H$ will have been determined.\cite{nlc}
As the Yukawa potential is much short-ranged compared to the toponium
resonance size, it does not affect the resonance spectra but makes the
wave function at the origin significantly larger.
Thus, only the normalization of the total cross section is changed,
while
the momentum distribution of top quark is quite insensitive to the
Yukawa coupling.
Fig.~8 demonstrates the dependence of the total cross section on
$M_H$ and $g_{tH}$.

If Higgs particle turns out to be heavier than $\sim 400$~GeV, it is
likely that a 500~GeV NLC cannot observe it.
In such a case, one may still be able to predict the Higgs mass through
loop effects.
Although the Higgs mass dependences of the loop effects are
small, using the interrelation among the precisely
determined top quark mass, the final data of LEP I,
the direct $W$ mass measurements, and left-right asymmetry
for polarized electron beams,
we may guess the next energy region to search for Higgs.
See Fig.~9.

\subsection{Forward-Backward Asymmetry of Top}

One of the unique features of toponium resonances is the
appearance of observable
FB-asymmetry of top quark in the threshold region.
The main contribution to the FB-asymmetry stems from the interference of
the vector
and axial-vector vertices.\cite{fb,nlo}

In the $e^+e^- \rightarrow t\bar{t}$ process, one can show from the spin-parity
argument that the $t\bar{t}V ~ (V=\gamma,Z)$ vector vertex creates S and D-wave
resonance states, while the $t\bar{t}Z$ axial-vector
vertex creates P-wave states.
Since the P-wave amplitude is suppressed by a
power of $\beta$ near threshold, its
interference with S-wave resonance states gives rise to the order
$\beta$ correction to the leading S-wave contribution to the cross
sections.  Since the interference term of the
vector and axial-vector couplings is proportional to $\cos
\theta$, S-P interference produces the FB-asymmetry at $O(\beta )$.

In general, S-wave resonance states and P-wave resonance states have
different energy spectra.
So if the c.m.\ energy is fixed at either of the spectrum, there would be no
contribution from the other.
However, the widths of resonances grow
rapidly as $m_t$ increases, and they become so large that S-wave and
P-wave resonance states start to interfere for $m_t \hbox
{ \raise3pt\hbox to 0pt{$>$}\raise-3pt\hbox{$\sim$} }$100~GeV.
This gives rise to the FB-asymmetry even below threshold, and provides
information on the resonance level structure which is concealed in the total
cross section due to the large smearing effect.

In Fig.~10 we show the FB-asymmetry as a function of energy measured
from the 1S resonance mass, $\Delta E = E - E_{1S}$.
One sees that the asymmetry increases up to
the 10\% level as the energy is raised.
This is because the energy levels of the resonances appear closer to
one another for larger $E$, and the overlap of the S and P-wave states
become more severe.
Note the $\alpha_s$-dependence of the FB-asymmetry:
the asymmetry decreases as $\alpha_s$ is raised.
This is because the resonance levels spread apart from each other with the
growth of binding energy, and also because the \oalfs \, correction of
the top quark width reduces the width for larger $\alpha_s$.

So, in essence, the FB-asymmetry `measures' the degree of overlap of
S-wave and P-wave resonances.

\subsection{Effects of Next-to-Leading Order Corrections}

As discussed in subsection 2.a, diagrams with the power counting
$\sim \alpha_s^{n+1}/\beta^n$ gives the order 10\% corrections to
leading order cross sections.
Recently, full next-to-leading order corrections have been calculated for
$\epem \to \ttbar$\cite{my,fkm2,nlo},
and also some of the second order corrections are
available.\cite{mk}

In addition to the corrections which were calculated
for bottomonium and charmonium,
there appear corrections that are relevant only to toponium.
For example, the
final state interaction diagrams which contribute at the
next-to-leading order
are shown in Fig.~11.
The contribution of the final state interaction
to the top quark momentum distribution can
be written as\cite{nlo}
\bea
\frac{d\sigma^{(FI)}}{d|\vc{p}|} \propto
\alpha_s \int \mbox{$\frac{d^3\vc{q}}{(2\pi )^3}$} \,
\frac{1}{|\vc{p} \! - \! \vc{q}|^3} \, Im \left[
\tilde{G}^*(\vc{q};E) \tilde{G}(\vc{p};E) \right] .
\eea
As the integrand on the right-hand-side changes sign for the
exchange of variables $\vc{p}$ and $\vc{q}$, the
left-hand-side vanishes upon integration over $\int d^3\vc{p}$.
Namely, it turns out that the corrections from the final state
interactions to the total cross section vanish\cite{my,fkm2,nlo}:
\bea
\int d\sigma^{(FI)} = 0 .
\eea

Nevertheless, final state interactions give non-trivial corrections
to the top quark momentum distributions.
Shown in Fig.~12 are the momentum distributions with
(solid) and without (dashed) the final state interactions.
The peak position of the distribution $|\vc{p}|_{peak}$ is reduced
about 5\% due to the correction for $m_t = 150$~GeV.
One sees that the final state interaction
correction grows for larger energy.
One can also show that this correction decreases as
$\Gamma_t \to 0$ as anticipated.

\section{Parameter Determination}

In this section we list the accuracy
which NLC has potential to achieve in the determination
of physical parameters
through the study of \ttbar
threshold cross sections.
The following are the results from the quantitative studies
for the sample case of $m_t = 170$~GeV,
taking into
account realistic experimental conditions expected at NLC.\footnote{
These are derived from the same analyses as done in Ref.\cite{fms}
where the case $m_t=150$~GeV is studied.
}
The error estimates include statistical ones only.

We can measure the threshold shape by energy scan.
Given 11 energy points with 1 fb$^{-1}$ each, $\Delta m_t = 380$~MeV is
expected even if $\alpha_s$ is totally unconstrained.
On the other hand, $\Delta \alpha_s(m_Z)=0.007$ even if $m_t$ is
unknown.
The error on top width is $\Delta \Gamma_t / \Gamma_t = 0.18$ when
$\alpha_s(m_Z)$ is known.
For the standard model Higgs of $M_H=100$~GeV, the Yukawa coupling can
be determined with $\Delta g_{tH} / g_{tH}=0.25$ if $\alpha_s$ and
$\Gamma_t$ are known from other sources.

The momentum spectrum measurement at \ttbar threshold can also be used.
Using the 1S peak position $E_{1S}$ determined by the threshold scan, we
can perform precision measurements of $\alpha_s(m_Z)$.
$\Delta \alpha_s(m_Z)=0.002$ is expected for 100 fb$^{-1}$ provided
$E_{1S}$ is known.
Given the same statistics,
$\Delta \Gamma_t / \Gamma_t = 0.03$ if $E_{1S}$ and $\alpha_s(m_Z)$
are known.
Note that since the top momentum measurement is insensitive to the
Higgs-exchange effect, the parameters determined from the top momentum
measurement can be fed back to the Yukawa coupling extraction from the
total cross section.

The FB-asymmetry of top quark gives additional information on $\alpha_s$
and $\Gamma_t$.
40k detected events provide us with an opportunity to measure
$\alpha_s(m_Z)$ with
$\Delta \alpha_s(m_Z)=0.004$ if $E_{1S}$ and $\Gamma_t$ are known.
Also,
$\Delta \Gamma_t / \Gamma_t = 0.06$ if $E_{1S}$ and $\alpha_s(m_Z)$
are known.

\section{Conclusion}

There appear many unique features in top quark threshold physics
as compared to lighter quarkonia.

Theoretical predictions of cross sections near \ttbar threshold are well
under control.
This is partly because we are reaching asymptotic region of QCD
potential, and partly because the large top width acts as infra-red
cutoff.
Full next-to-leading order corrections as well as part
of the second order corrections are already available.

The \ttbar threshold cross sections depends on physical parameters such
as $m_t$, $\alpha_s$, $\Gamma_t$, $M_H$, and $g_{tH}$.
Essentially there are three independent observables that are unique to
the \ttbar threshold region.
The total cross section is sensitive to the short distance physics
such as Higgs-exchange effect.
The top quark momentum measurement will allow to extract information on
the resonance wave function, and it turns out to be most sensitive to
$\alpha_s$ and $\Gamma_t$.
The FB-asymmetry of top quark in the threshold region 'measures' the
level gaps between S and P-wave resonance states, and provides another
information on the physical parameters.

Although not covered in this paper, one may also perform other standard
top quark measurements (top form factors, exotic decays, etc.)\
which can be done both at threshold and in open
top region ($E \gg 2m_t$).\cite{nlc,jez}
There seems to be some advantage of the threshold region, since the cross
section is largest, and also since highly polarized top quark
samples can be obtained.
These remain for future study.

In conclusion, the top quark threshold is quite rich in physics, and
deserves serious studies.

\section*{Acknowledgements}

Most of the topics covered here are based on top quark studies
in collaboration with K. Fujii, K. Hagiwara, K.
Hikasa, S. Ishihara, T. Matsui, H. Murayama and C.-K. Ng.
The author wishes to express his sincere gratitude to all of them.
The quantitative re-analysis for $m_t=170$~GeV case
given in section 4 has been done by K. Fujii.
The author is also grateful to
M.~Je\.{z}abek, J.~K\"{u}hn, K.~Melnikov, W.~M\"{o}dritsch,
M.~Peskin, T. Teubner,
O.~Yakovlev,
and all the members of JLC Working Group
for fruitful discussions.
Finally, the author wishes to thank T. Moroi and T. Asaka for reading
the manuscript carefully and making useful comments.



\newpage
\section*{Figure Captions}
\renewcommand{\labelenumi} {Fig. \arabic{enumi}}
\begin{enumerate}

\item 

Top quark pair production near threshold. Top quark
will decay before hadronization occurs.

\item 

The ladder diagrams for the process $\gamma^* \to t\bar{t}$.  The
diagram where $n$ uncrossed gluons are exchanged has the behavior $\sim
(\alpha_s/\beta)^n$ near threshold.

\item 

The Cutkosky rule for evaluating the imaginary part of the 1-loop
diagram.  The factors in $\alpha_s$ and $\beta$ are shown explicitly.

\item 

The cut-diagram method for evaluating the singularities of the
higher order ladder diagrams.  The factors in $\alpha_s$ and $\beta$ are
shown explicitly.

\item 

The self-consistent equation satisfied by the leading singularities
of the $t\bar{t}\gamma$ vertex $\Gamma^\mu_0$.  One should take only the
leading part $\sim (\alpha_s/\beta)^n$ on both sides of the equation.

\item 

(a) An example of energy scan to determine $m_t$ and $\alpha_s(m_Z)$
where each point corresponds to 1~fb$^{-1}$.
(b) The contour resulting from the fit to the data points.\cite{fujii}

\item 

The top momentum distribution at the c.m.\ energy
2~GeV above the first peak, $\Delta E =
E - E_{1S} = 2$~GeV, for $\alpha_s(m_Z) = 0.11$, 0.12, and 0.13 with
$m_t = 170$~GeV.\cite{fujii}

\item 

An example of energy scan to determine the Higgs mass ($M_H$) and the
normalized top-Higgs Yukawa coupling ($\beta_H =
g_{tH}/g_{tH}^{SM}$)
where the cross section curves are superimposed for several $M_H$
values.
(b) The same plot with the cross section curves for several $\beta_H^2$
values.
(c) The contour resulting from the fit to the data points.\cite{fujii}

\item 

(a) The 1-$\sigma$ contours in the $m_t$-$M_H$ plane for $\Delta M_W =
21$~MeV and $\Delta m_t = 0.5$, 5.0, and 10.0~GeV.
The bounds on Higgs mass quickly shrink with the increasing
accuracy of $m_t$.
Input values are $m_t=170$~GeV and $M_H=500$~GeV.
(b) The 1-$\sigma$ bounds on the Higgs mass as a function of the input
Higgs mass when we assume $\Delta M_W=21$~MeV and
$\Delta m_t=0.5$~GeV.\cite{fujii2}

\item 

The forward-backward asymmetry of top quark
as a function of the energy measured from
the first peak, $\Delta E = E - E_{1S}$, showing the dependence on
(a) $\alpha_s(m_Z)$, (b) $|V_{tb}|^2 = \Gamma_t/\Gamma_t^{SM}$, and
(c) $m_t$.\cite{fujii}

\item 

Diagrams representing the final state interactions, where the gluon is
exchanged between $t(\bar{t})$ and $\bar{b}(b)$ and between $b$ and
$\bar{b}$.

\item 

The top momentum distribution
$|\vc{p}|^2 T_0$ versus $|\vc{p}|$ with
(solid) and without
(dashed) the
final state interaction corrections for $m_t = 150$~GeV and
$\alpha_s (m_Z) = 0.12$.\cite{nlo}

\end{enumerate}

\end{document}